\documentclass{article}
\usepackage{spconf,amsmath,graphicx}
\usepackage{amssymb}
\usepackage{mwe} 
\usepackage{booktabs}
\usepackage{amsfonts}
\usepackage{siunitx}
\usepackage{xspace}
\usepackage{xurl} 

\title{Multimodal MRI Report Findings Supervised Brain Lesion \\ Segmentation with Substructures}

\name{\parbox{\textwidth}{\centering 
    Yubin Ge$^{1\dag}$,  Yongsong Huang$^{2,3\dag}$,  
    Xiaofeng Liu$^{2*}$
}}

\address{$^{1}$ Amazon AWS, 
$^{2}$ Yale University, $^{3}$ Tohoku University}

\begin{document}
 
\maketitle

\begingroup
  \renewcommand\thefootnote{} 
  \footnotetext{$^{\dag}$ contribute equally. $^{*}$corresponding author:  {xiaofeng.liu@yale.edu}}
\endgroup

\begin{abstract}
Report-supervised (RSuper) learning seeks to alleviate the need for dense tumor voxel labels with constraints derived from radiology reports (e.g., volumes, counts, sizes, locations). In MRI studies of brain tumors, however, we often involve multi-parametric scans and substructures. Here, fine-grained modality/parameter-wise reports are usually provided along with global findings and are correlated with different substructures. Moreover, the reports often describe only the largest lesion and provide qualitative or uncertain cues (``mild,'' ``possible''). Classical RSuper losses (e.g., sum volume consistency) can over-constrain or hallucinate unreported findings under such incompleteness, and are unable to utilize these hierarchical findings or exploit the priors of varied lesion types in a merged dataset. We explicitly parse the global quantitative and modality-wise qualitative findings and introduce a unified, one-sided, uncertainty-aware formulation (MS-RSuper) that: (i) aligns modality-specific qualitative cues (e.g., T1c enhancement, FLAIR edema) with their corresponding substructures using existence and absence losses; (ii) enforces one-sided lower-bounds for partial quantitative cues (e.g., largest lesion size, minimal multiplicity); and (iii) adds extra- vs.\ intra-axial anatomical priors to respect cohort differences. Certainty tokens scale penalties; missing cues are down-weighted. On 1238 report-labeled BraTS-MET/MEN scans, our MS-RSuper largely outperforms both a sparsely-supervised baseline and a naive RSuper method.
\end{abstract}


\begin{keywords}
Report supervision, multimodal MRI, meningioma, brain metastases, segmentation.
\end{keywords}

\vspace{-2pt}
\section{Introduction}
Accurate delineation of lesion structures is a fundamental step in clinical diagnosis, intervention, and treatment planning~\cite{liu2023incremental}. However, voxel-wise annotation for 3D multimodal MRI is costly and subjective, especially when substructures of tumor core (TC), enhancing tumor (ET), and edema (ED) must be delineated across sequences with different contrast mechanisms~\cite{liu2023incremental}. {How to utilize existing, routinely summarized radiology reports to aid (or assist) segmentation model training is of great importance} for more practical utilization of big medical data.

To exploit the valuable and relatively large-scale text information, early attempts usually form a multi-task learning with an auxiliary task in addition to the segmentation with either the extracted tumor present/absent label for classification~\cite{zhang20213d}, or a contrastive language-image pre-training objective~\cite{blankemeier2024merlin}. {While the benefits of auxiliary tasks to segmentation are indirect and occasionally minor, the recently developed} Report-supervised learning (RSuper)~\cite{bassi2025learning} promises to directly leverage the detailed volumes, counts, and sizes extracted from abundant abdominal CT radiology reports and enforces the corresponding loss functions alongside the conventional segmentation loss for only a small portion of segmentation labeled samples (e.g., 50 scans). {This approach therefore largely reduces the burden of manual labeling.}

However, in MRI studies of brain tumors, we often involve multi-parametric scans of T1, T1c, T2, and FLAIR as well as substructures of TC, ET, and ED. In which the fine-grained modality/ parameter-wise finding reports are usually provided along with the global finding reports, and correlated to different substructures. How to systematically exploit both the \emph{global} descriptors (falx-/skull-base adjacency vs.\ deep parenchyma, approximate size, multiplicity, edema/midline shift) and  \emph{modality-wise} descriptors (T1c enhancement pattern; FLAIR hyperintensity) is largely underexplored. 

Moreover, the reports often describe only the largest lesion, provide qualitative or uncertain cues (“mild,” “possible”). Specifically, in brain lesions, reports are often \emph{partially specified}: many cases provide only the \emph{largest} lesion size $d_{\max}$, omit axes for diameters, and/or use certainty qualifiers (\emph{mild}, \emph{possible}, \emph{equivocal}); and the counts are sometimes qualitative (``multiple''). Simply adopt the classical RSuper volume loss, which penalizes the difference of predicted and labeled sum volume of all tumors, may (a) learn to suppress small lesions that the report does not enumerate, or (b) \emph{unduly shrink} tumors to match a partial volume hint.

Finally, when combining image-report records from multiple diseases, cohort-specific priors are lost. For example, \emph{BraTS-MET} (metastases) typically includes multiple intra-axial parenchymal lesions, often with ring enhancement. In contrast, \emph{BraTS-MEN} (meningioma) is typically extra-axial and dural-based (e.g., falx, skull base) with solid enhancement. A report-derived supervision should transfer across cohorts without imposing contradictory biases. A naive RSuper loss cannot leverage these strong anatomical priors.

To address these limitations, we propose a novel multimodality with substructure RSuper framework for brain lesion segmentation. Our main contributions are:
 
$\bullet$\textbf{Modality-Substructure Alignment:} We introduce a loss that links modality-specific report findings (e.g., T1c enhancement, FLAIR edema) directly to their corresponding segmentation substructures (ET and ED, respectively).

$\bullet$\textbf{One-Sided Partial-Report Loss:} We propose a "lower-bound" size loss and "minimal-multiplicity" count loss to handle incomplete reports that only describe the largest lesion or use qualitative counts, avoiding penalty for valid, unreported lesions.

$\bullet$\textbf{Cohort-Specific Priors:} We integrate an anatomical prior loss that penalizes intra-axial predictions for MEN and extra-axial predictions for MET, guided by cohort-level cues from the reports.
 
We validated its effectiveness on the combined BraTS-MET and BraTS-MEN with segmentation and report\footnote{\url{https://huggingface.co/datasets/JiayuLei/RadGenome-Brain_MRI/tree/main}}.

\vspace{-1pt}
\section{Methodology}\vspace{-1pt}
\label{sec:method}

Our framework trains a 3D segmentation network using a partially segmented dataset, i.e., a large set of image-report pairs ($D_R$) and a small set of fully-masked data ($D_M$). The model is trained with a composite loss. For data in $D_M$, we use a standard supervised segmentation loss, $L_{\text{seg}}$ (e.g., a combination of Dice and Cross-Entropy loss). For data in $D_R$, we introduce a novel report-supervised loss, $L_{\text{report}}$, designed to handle the hierarchical, qualitative, and partial nature of multimodal MRI reports.

\vspace{-1pt}
\subsection{Hierarchical Report Parsing and Mapping}\vspace{-1pt}
\label{sec:parsing}

We first employ a Large Language Model (LLM) with domain-specific prompts to parse each free-text radiology report. Critically, we categorize the extracted cues into two distinct types based on their nature and scope:

\noindent \textbf{(A)} \textbf{Quantitative Global Cues:} These are specific measurements, typically found in the "global findings" section, that apply to the \emph{entire} lesion or provide a total count. These are often partial (e.g., ``largest lesion measuring 45x39x47 mm," "multiple punctate... lesions").

\noindent \textbf{(B)} \textbf{Qualitative Modality-Specific Cues:} These are usually descriptive, non-numeric findings tied to a specific MRI sequence, which inherently map to tumor substructures. Though sometimes the tumor size is provided, it is the same as the global finding, which does not provide incremental information. For example, {T1c} often includes ``obvious enhancement," ``ring enhancement," ``no enhancement," while {FLAIR} includes ``surrounding extensive edema," ``mild hyperintense signal."
 
Based on this parsing, we propose to establish a \textbf{Modality-Substructure Alignment Principle}. This is not a loss function itself, but a crucial mapping rule that directs how constraints are applied:
 
$\bullet$ T1c findings (enhancement) constrain the {Enhancing Tumor ($P_{ET}$)} probability map.

$\bullet$ FLAIR findings (edema, non-specific hyperintensity) constrain the {Edema ($P_{ED}$)} probability map.

$\bullet$ T1 or T2 findings (e.g., "hypointense core") constrain the {Tumor Core ($P_{TC}$)} map.

$\bullet$ Global cues (e.g., total size, count) constrain the {Whole Tumor ($P_{WT}$)} map, where $P_{WT} = P_{ET} + P_{ED} + P_{TC}$.
 
Uncertainty cues (e.g., "possible," "mild") are parsed into a scaling weight $\lambda \in [0, 1]$ for the corresponding loss term.

\begin{figure}[t]
    \centering
    \includegraphics[width=1\linewidth]{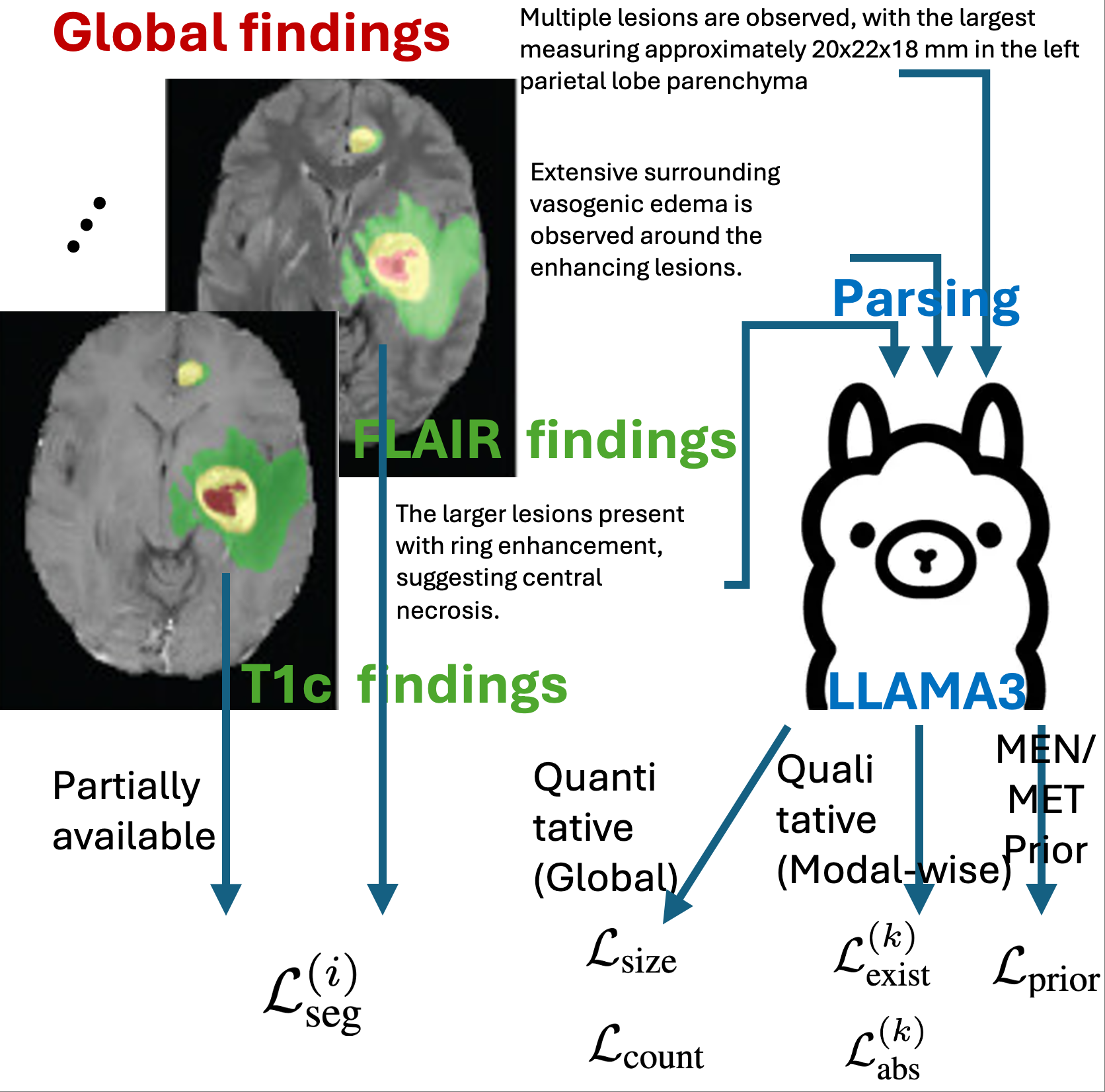} \vspace{-4mm}
    \caption{Overview of our proposed report-supervised framework. An LLM parses hierarchical findings from reports. Our losses align modality-specific findings, handle partial cues, and enforce anatomical location priors.}
    \label{fig:framework}
    \vspace{-3mm} 
\end{figure} 

\vspace{-3pt}
\subsection{Unified Report Constraint Loss ($\mathcal{L}_{\text{report}}$)}\vspace{-1pt}
\label{sec:constraint_loss}

Our primary report loss, $\mathcal{L}_{\text{report}}$, combines constraints from both qualitative and quantitative cues, applying them to the aligned substructure maps identified in $\S 2.1$.
\vspace{-3pt}

\subsubsection{Substructure Qualitative Existence and Absence Loss}
As identified, most modality-specific cues are qualitative (e.g., "edema is present") and lack quantitative volumes. We cannot use a symmetric L1/L2 volume loss~\cite{bassi2025learning}. Instead, we formulate a loss based on the \emph{existence} or \emph{absence} of a finding. For a given substructure class $k$ (e.g., $k \in \{\text{ET, ED, TC}\}$), let $V_{k}$ be the predicted volume of $P_k(\mathbf{x})\geq0.5$ for that substructure.

If the report confirms the \textbf{presence} of substructure $k$ (e.g., "surrounding edema" $\rightarrow k=\text{ED}$) with confidence $\lambda_{k, \text{pos}}$, we apply an "Existence Loss." This loss penalizes the model only if it fails to predict any presence (volume $<1$) of that substructure $\mathcal{L}_{\text{exist}}^{(k)} = \max(0, 1- V_{k}),$
which encourages the model to segment at least 1 voxel for class $k$, without hallucinating a specific target volume.

Conversely, if the report explicitly confirms the \textbf{absence} of a substructure $k$ (e.g., "no enhancement" $\rightarrow k=\text{ET}$), we apply an loss that penalizes any prediction for that class $
\mathcal{L}_{\text{exist}}^{(k)} =  V_{k}.$ Therefore, we have 
\begin{equation}
\mathcal{L}_{\text{exist}}^{(k)} = 
\begin{cases} 
  \max(0, 1 - V_{k}) & \text{if presence confirmed,} \\
  V_{k} & \text{if absence confirmed,} \\
0 & \text{otherwise.}
\end{cases}
\label{eq:qual_loss}
\end{equation}

\vspace{-1pt}
\subsubsection{Global One-Sided Partial Cue Loss (Size and Count)}\vspace{-1pt}
We handle the quantitative but partial nature of global cues:

 $\bullet$ \textbf{Size Loss:} Reports often provide only the 3D-dim or diameter(s) of the \emph{largest} lesion, $d_{\max}$. Let $C_{\text{pred}}$ be the set of predicted connected components for the whole tumor ($P_{WT} \geq 0.5$). Let $d_c$ be the volume of a component $c \in C_{\text{pred}}$. The loss is:
$$
\mathcal{L}_{\text{size}} = |d_{\max} - \max_{c \in C_{\text{pred}}} d_c|.
$$
This loss used mean absolute error, which is more robust to small inaccurate measure of $d_{\max}$.


$\bullet$ \textbf{Count Loss:} Reports often use qualitative counts like "multiple" or "a few." We parse this to a minimal integer $N_{\text{qual}}$ (e.g., "multiple" $\rightarrow N_{\text{qual}} = 2$). We apply a one-sided count loss:
$$
\mathcal{L}_{\text{count}} = \max(0, N_{\text{qual}} - |C_{\text{pred}}|).
$$
It penalizes the model if it predicts \emph{fewer} than $N_{\text{qual}}$ lesions.

Therefore, we have$\mathcal{L}_{\text{global}}=w_{\text{size}} \mathcal{L}_{\text{size}} +  w_{\text{count}}\mathcal{L}_{\text{count}}$.

\vspace{-1pt}
\subsection{Cohort-Specific Anatomical Prior Loss ($\mathcal{L}_{\text{prior}}$)}\vspace{-1pt}
\label{sec:prior_loss}
Finally, we leverage global location cues (e.g., ``falx," ``parenchymal") to identify the cohort (MEN or MET) and apply a strong anatomical prior. We use pre-defined binary masks for the dura/extra-axial space ($M_{\text{dural}}$) and the brain parenchyma/ intra-axial space ($M_{\text{parench}}$).

$\bullet$ If the report suggests \textbf{Meningioma (MEN)}, which is extra-axial, we penalize any intra-axial predictions:
    $$
    \mathcal{L}_{\text{prior}} =  \sum_{\mathbf{x}} (P_{WT}(\mathbf{x}) \cdot M_{\text{parench}}(\mathbf{x})).\vspace{-3pt}
    $$
    
$\bullet$ If the report suggests \textbf{Metastases (MET)}, which are intra-axial, we penalize extra-axial predictions:
    $$
    \mathcal{L}_{\text{prior}} =  \sum_{\mathbf{x}} (P_{WT}(\mathbf{x}) \cdot M_{\text{dural}}(\mathbf{x})).\vspace{-3pt}
    $$
This loss effectively guides the model to search in the correct anatomical compartment, resolving ambiguity and reducing false positives.

\vspace{-1pt}
\subsection{Total Loss Function}\vspace{-1pt}
The model is first pre-trained on $D_M$ using $L_{\text{seg}}$. It is then fine-tuned on the combined dataset $D_M \cup D_R$. For a mixed batch $B = B_M \cup B_R$, the total loss is:
$$
\mathcal{L}_{\text{total}} = \frac{1}{|B_M|} \sum_{i \in B_M} \mathcal{L}_{\text{seg}}^{(i)} + \frac{w_r}{|B_R|} \sum_{j \in B_R} \mathcal{L}_{\text{report}}^{(j)},
$$
where $w_r$ is a balancing weight for the report-based supervision, and $\mathcal{L}_{\text{report}}$ is the sum of our proposed constraint losses:
\begin{align*}
\mathcal{L}_{\text{report}} = \sum_{k}^3 \mathcal{L}_{\text{exist}}^{(k)}  + w_{\text{size}} \mathcal{L}_{\text{size}} +  w_{\text{count}}\mathcal{L}_{\text{count}} + w_{\text{prior}} \mathcal{L}_{\text{prior}},
\end{align*}
where $w_{\text{size}}$, $w_{\text{count}}$, and $w_{\text{prior}}$ are weights for each component of the report-supervised loss.

\section{Experiments and Results}\vspace{-3pt}
\label{sec:exp}

We used two large-scale, multi-modal MRI segmentation datasets. Their associated radiology reports are manually generated in 
RadGenome-Brain\_MRI dataset~\cite{lei2024autorg}$^1$. Each subject has a global finding and four modality-wise findings.
 
$\bullet$ \textbf{BraTS-MEN (Meningioma):} A collection of 1000 subjects (4000 3D mpMRI scans) with meningioma. Reports frequently describe extra-axial, dural-based lesions (e.g., "falx cerebri," "skull base," "cerebellopontine angle") with "marked," "uniform" enhancement on T1c.

$\bullet$ \textbf{BraTS-MET (Metastases):} A collection of 238 subjects (952 3D mpMRI scans) with brain metastases. Reports describe "multiple," "parenchymal" (intra-axial) lesions, often with "ring enhancement" and "extensive surrounding edema" on FLAIR.
 

We held out 50 MEN and 50 MET for testing, and remaining subjects for training (all with reports, while 50 MEN and 50 MET has segmentation masks). The LLM parser was implemented using Llama 3.1 70B as in~\cite{bassi2025learning} with prompts engineered to extract the hierarchical attributes, uncertainty weights, and cohort priors.


We used the 3D nnU-Net framework as our base segmentation architecture due to its strong performance. The model was pre-trained on BraTS2018 for Glioblastoma (different tumor from MEN or MET) with the supervised CE loss for labeled substructures of TC, ET, and ED~\cite{liu2023incremental}. No report is available. Loss weights were set empirically as $w_r=0.2$, $w_{\text{size}}=1.0$, $w_{\text{count}}=0.5$, and $w_{\text{prior}}=0.2$.

We compare three methods: (1) \textbf{Masks-Only (Baseline):} {fine-tuned} only on the 100 labeled scans ($D_M$). (2) \textbf{R-Super}~\cite{bassi2025learning} finetuned on $D_M \cup D_R$, using the summed volume and count applied to the "Whole Tumor" (WT = ET+ED+TC) prediction. (3) \textbf{Ours MS-RSuper:} multimodal with substructure supervised by $\mathcal{L}_{\text{report}}$.


As shown in Table \ref{tab:main_results}, our method largely outperforms both baselines. The Masks-Only model suffers from poor generalization, as expected from only 50 labels in each disease. The RSuper \cite{bassi2025learning} baseline dose not provides improvement, since its symmetric ``summed volume" loss is confused by the partial reports (e.g., only the largest lesion), leading to suboptimal performance. Our MS-RSuper achieves the highest Dice scores across all substructures and both cohorts. The gains are promising in the MET dataset, where our $\mathcal{L}_{\text{count}}$ (handling ``multiple" lesions) and our qualitative losses ($\mathcal{L}_{\text{exist}}$) (which align 'edema' to $P_{ED}$ and `enhancement' to $P_{ET}$) are critical. For the MEN dataset, $\mathcal{L}_{\text{prior}}$ (enforcing extra-axial location) was key to reducing false positives in the brain parenchyma.

\begin{table}[t]
\centering
\caption{Dice Score on the held-out test sets.}
\label{tab:main_results}
\resizebox{\linewidth}{!}{%
\begin{tabular}{@{}llccc@{}}
\toprule
\textbf{Method} & \textbf{Test Set} & \textbf{WT (DSC)} & \textbf{TC (DSC)} & \textbf{ET (DSC)} \\ \midrule
Masks-Only  & MEN & 0.481 & 0.323 & 0.370 \\
R-Super \cite{bassi2025learning} & MEN & 0.452 & 0.301 & 0.353 \\
\textbf{MS-RSuper(Ours)} & MEN & \textbf{0.554} & \textbf{0.428} & \textbf{0.489} \\ \midrule
Masks-Only  & MET & 0.420 & 0.385 & 0.321 \\
RSuper \cite{bassi2025learning} & MET & 0.443 & 0.391 & 0.333 \\
\textbf{MS-RSuper(Ours)} & MET & \textbf{0.529} & \textbf{0.494} & \textbf{0.452} \\ \bottomrule
\end{tabular}%
}
\vspace{-4mm} 
\end{table}

An ablation study (Table \ref{tab:ablation}) on the MET dataset confirms that each of our proposed loss components contributes to the final performance. The quantitative, partial-cue losses ($\mathcal{L}_{\text{exist}}$) provide the first major boost by handling size and count. Adding the qualitative modality-aligned losses ($\mathcal{L}_{\text{global}}$) further improves performance by correctly using the T1c and FLAIR cues. Finally, the cohort prior ($\mathcal{L}_{\text{prior}}$) provides an additional gain by penalizing anatomically implausible predictions.



\vspace{-8pt}
\section{Conclusion}\vspace{-3pt}
\label{sec:conclusion}

We introduced a novel report-supervised learning framework tailored for the complexities of multi-parametric brain MRI and substructure segmentation. Unlike prior work on CT, our method addresses three key challenges: (1) it aligns qualitative modality-specific findings with their corresponding segmentation substructures using novel existence and absence losses, (2) it uses one-sided, uncertainty-aware losses to robustly handle partial quantitative reports (e.g., ``largest lesion only," ``multiple"), and (3) it integrates cohort-level anatomical priors (intra- vs. extra-axial) derived from report keywords. We evaluate on a large dataset of 1238 meningioma and metastases scans, our approach largely outperformed both a sparsely-supervised baseline and a naive application of existing RSuper methods. This work demonstrates that by designing losses that faithfully reflect the hierarchical and often-incomplete nature of radiology reports, we can effectively leverage large-scale text data to improve multi-class segmentation in multimodal imaging.

\begin{table}[t]
\centering
\caption{Ablation study on the BraTS-MET Test set (WT Dice). Each component of our proposed loss ($\mathcal{L}_{\text{exist}}$, $\mathcal{L}_{\text{global}}$, $\mathcal{L}_{\text{prior}}$) provides a cumulative benefit.}
\label{tab:ablation}
\resizebox{0.8\linewidth}{!}{%
\begin{tabular}{@{}lc@{}}
\toprule
\textbf{Method} & \textbf{WT (DSC)} \\ \midrule
Masks-Only (Baseline) & 0.420 \\
+ $\mathcal{L}_{\text{exist}}$ (Partial size/count) & 0.475 \\
+ $\mathcal{L}_{\text{exist}} + \mathcal{L}_{\text{global}}$ (Adds qualitative) & 0.513 \\
+ $\mathcal{L}_{\text{exist}} + \mathcal{L}_{\text{global}} + \mathcal{L}_{\text{prior}}$ (Full MS-RSuper) & \textbf{0.529} \\ \bottomrule
\end{tabular}%
}
\vspace{-3mm} 
\end{table}

\vspace{-2pt}
\section{COMPLIANCE WITH ETHICAL STANDARDS}\vspace{-3pt}
This retrospective study used open-access human subject data; no additional ethical approval was required.

\vspace{-2pt}
\section{ACKNOWLEDGMENTS}\vspace{-3pt}
Supported in part by NIH R21EB034911 and NVIDIA Academic Grant Program.

 \vspace{-3pt}
 
\bibliographystyle{IEEEbib}
\bibliography{refs}
\end{document}